# Order within disorder: spectral key generation and distribution in random lasers

*Zhijia Hu[1,*], Shilong He[1], Lianghao Qi[1], Yalan Li[1], Siqi Li[1,*], Bin Chen[3], Wenyu Du[1], Yan Kuai[1], Zhigang Cao[1], Min Wang[1], Kaiming Zhou[4], Lin Zhang[4], Qingchuan Guo[5], Weimin Ding[6], Chao Li[7], Kang Xie[8], Anderson S.L. Gomes[2], Benli Yu[1]*

[1]Information Materials and Intelligent Sensing Laboratory of Anhui Province, National Key Laboratory of Opto-Electronic Information Acquisition and Protection Technology, School of Physics and Opto-electronics Engineering, Anhui University, Hefei 230601, China

[2]Departamento de Física, Universidade Federal de Pernambuco, 50670-901, Recife, PE, Brazil

[3]School of Computer and Information Engineering, Hefei University of Technology (HFUT), Hefei 230601, China

[4]Aston Institute of Photonic Technologies, Aston University, Birmingham B4 7ET, United Kingdom

[5]Center for Lasers and Optics, Anhui University, Hefei 230088, China

[6]China Mobile Group Design Institute Co., Ltd. Anhui Branch, Hefei 230601, China

[7]Peng Cheng Laboratory, Shenzhen 518055, China

[8]State Key Laboratory of Precision Electronic Manufacturing Technology and Equipment, School of Electromechanical Engineering, Guangdong University of Technology, Guangzhou 510006, China

[*]e-mail: zhijiahu@ahu.edu.cn, sqli@ahu.edu.cn

In secure communication, highly random entropy sources are essential for information security. Random lasers (RLs), which arise from multiple scattering in disordered structures, are potentially ideal entropy sources. Traditionally, RLs are viewed as disordered and unpredictable. However, in this work, we present novel evidence that orderly patterns exist beneath the seemingly disordered outputs of RLs. Utilizing deep learning techniques, a variety of advanced neural network models are used to analyze the spectral data in multiple dimensions. The results show that the time series of RLs spectra are unpredictable, but spectral wavelength component intensities can be recovered due to inter-modal correlations. This finding not only breaks through the traditional perception that RLs are unpredictable, but also reveals for the first time that RLs have the dual characteristics of both randomness and determinism. Based on this new characteristic, we further expand the application field of RLs and innovatively design a new type of key generation and distribution scheme. In this scheme, the disordered property of RLs is used for key generation to ensure high randomness, while their ordered property is used for key distribution to guarantee accuracy and reliability. The scheme provides a new strategy for secure communication.

**1. Introduction**

In today's information society, the demand for protecting information is increasing. For most secure communication technologies, secure key generation and distribution (SKGD) determines the security of messages. Traditional key generation is based on mathematical algorithms. The generated pseudo-random sequences (PRS) are inevitably periodic and threatened by the evolution of the quantum computing paradigm [1]. For key transmission, SKGD is solved at the upper layer of the protocol stack through public key encryption algorithms that rely heavily on computational complexity as a security mechanism [1]. However, with the development of advanced quantum computers, the utility of these algorithms faces a real challenge [3][4]. In recent years, physical entropy sources such as electronic noise [5], chaotic circuits [6], and photonic entropy sources [7][8] have received much attention. The randomness of physical entropy sources originates from natural physical processes, which makes them difficult to predict and replicate with higher security. In addition, physical entropy sources are widely used in various key distribution schemes. The high randomness of entropy sources can directly affect the security of key distribution. The reported classical physical key distribution methods are mainly based on physical unclonable functions [9], fiber lasers [10], fiber channel noise [11], and electric or optical chaos [12][13]. However, there are many challenges in terms of security, efficiency and feasibility. In contrast, quantum key distribution promises unconditional security in principle, but its high cost, complexity, and critical rate distance limitation are still remaining challenges [14][15]. Therefore, SKGD methods based on physical entropy sources still need to be continuously optimized and innovated to meet the increasing information security requirements.

Recently, random lasers (RLs) have received much attention as optical random number generators [16][17]. RLs do not have a traditional cavity structure and require no reflective elements [18]. Photons are amplified by random multiple scattering in the gain medium to provide sufficient feedback for laser emission [19]. Due to the random scattering feedback, the number and phase of photons in the amplification process are also random. RL is highly random and is a potentially physical entropy source [20][21]. Indeed, theoretical models that describe the intensity distribution usually include additive and multiplicative noise sources in stochastic differential equation approaches [22][23]. RL can be obtained in many random media, including semiconductor powders, polymers, organic films and even biological tissues [24]-[27]. Among these random media, polymer optical fiber (POF) is widely used in short-distance optical communications, optical sensors, and other fields due to its good flexibility, ease of handling, economy, small core size, and large numerical aperture [28][29]. In a disordered medium, photons undergo multiple scatterings back to the initial scattering point, forming a closed loop that produces coherent RL radiation [30]. The radiation spectrum has discrete multi-spikes, called coherent feedback. Compared with incoherent RLs, coherent RLs exhibit richer spectral dynamics [30][31]. Spectra, as an important output characterization result of the randomness of random lasers, are rich in physical connotations. For a long time, RL has been considered as an unpredictable light source [32]-[34]. However, recent studies have revealed the observation of replica-symmetry-breaking phenomenon in the glass phase of RLs [35], which is centered on correlations between modes and competing nonlinear interactions [36][37]. The characterization of this phenomenon relies heavily on the analysis of Parisi's overlap distribution of replica correlations [38][39]. The inter-mode correlations imply that there may be certain deterministic patterns among the modes of the spectrum of a random laser.

   Deep learning as a technique for extracting hidden features has grown significantly in recent years. Contrary to shallow learning, deep learning usually refers to stacking multiple layers of neural networks and relying on random optimization for machine learning tasks [40]. In various models of deep learning, the multilayer perceptual machine (MLP) is the most basic structure, which consists only of fully connected layers, i.e., each neuron is connected to all neurons in the next layer [41]. Convolutional neural networks (CNNs) can convolutionally abstract features of the input data and preserve the relationships between image pixels [42]. Long short-term memory (LSTM) has the ability to memorize historical sequences and can store long sequences of information in hidden memory. LSTM is also continuously updated over time to ensure the continuity of temporal information [43]. The gated recurrent unit (GRU) is a simplified version of the LSTM, also used for processing sequential data [44]. The attention mechanisms can adaptively distinguish between different levels of important information and pay more attention to relevant and beneficial information [45]. The Transformer architecture, on the other hand, completely abandons the loop structure and is constructed based on the self-attention mechanism [46]. The Transformer model, which processes sequential data by means of multiple heads of attention and positional encoding, has achieved great success in the field of natural language processing. In recent years, many researchers have combined these deep neural networks (DNNs) models to extract temporal and spatial features in many domains, such as image captioning and time series prediction [47][48]. Some seemingly irrelevant chaotic sequences can be successfully predicted by DNNs due to the existence of fixed formulas [50][51], which undoubtedly brings a challenge to coherent RLs due to the existence of inter-modal correlations. In addition, deep learning is able to realize information recovery by virtue of its powerful learning ability, and has been gradually applied to the field of key distribution in recent years [52]. Deep learning is a powerful tool for abstracting potential features, which provides the possibility of researching the spectral potential features of coherent RLs. To the best of our knowledge, deep learning-based spectral characterization of coherent RL has not been reported previously.

Here, we demonstrate for the first time that coherent feedback RLs with random feedback and inter-modal correlation can be used to design secure key generation and distribution schemes. POF doped with laser dye and nanoparticles is used as a sample. The spectral data of the output coherent RL is analyzed in the time and frequency domains. The results show temporal intensity is irrelevant for separating wavelength channels, but correlation exists between wavelength components. The randomness of the coherent RL time series is tested by multiple state-of-the-art time series prediction models (e.g., MLP-Attention, GRU, CNN-LSTM-Attention, Transformer). The results show that the existing models are unable to predict its outputs. Coherent RL spectra are highly randomized in the time domain, which is expected to be used for secure key generation. In addition, we perceive that RLs may have some potential correlation between spectral components. The nonlinear contribution of each wavelength component is extracted by DNN model. High accuracy prediction of wave intensity is achieved (up to 99% accuracy for single wave prediction), revealing the determinism between its spectral components. This indicates that coherent RL is both stochastic and deterministic. Based on these properties, we design an innovative SKGD scheme. The communicating parties use the spectral wave intensity as the key. The transmitting end extracts the specific wave intensity and converts it to a key to complete the key generation. The remaining spectrum is transmitted to the receiving end. The receiving end recovers the wave peak information with the pre-trained DNN and converts it to the key to complete the key distribution. The results show that the prediction accuracy and stability are high and have certain anti-attack ability. The scheme not only generates the key securely, but also realizes the secure and efficient distribution of the key at the same time, which provides a new strategy in the field of secure communication.

## 2. Coherent random laser (RL) based on polymer optical fiber (POF)

### 2.1. Output properties

The experimental setup is shown in **Figure 1a**. POF doped with laser dye and nanoparticles is used as a sample. A 532 nm pulsed laser provides the pump energy, and a Glan-Prism group is used to regulate the pump energy. The beam splitter divides the pump light equally into two beams, one focused on the sample to excite a coherent feedback RL. A spectrometer collects the spectral data of the emission. The other beam is connected to an energy meter to synchronize the recording of the pump's energy. Fig. 1b shows the principle of RL generation. The pumped photons are subjected to multiple reflections from the nanoparticles and amplification gain from the PM597 laser dye. The presence of a closed loop allows for coherent feedback, enabling coherent RL emission. The one-dimensional confinement of the POF makes the coherent RL emission highly directional. Fig. 1c demonstrates the emission spectra of the coherent RL at different pump energies, where multimode is the characteristic of the RL rather than single mode due to multiple scattering. When the pump energy is less than 48 μJ, only a broad spontaneous emission spectrum centered at 572 nm is observed. As the pump energy exceeds 48 μJ, several discrete narrow peaks with line widths of about 0.5 nm are clearly observed, indicating that coherent resonant feedback is established in the sample to realize coherent RL emission. Fig. 1d demonstrates the variation of the emission intensity and line width of the RL at different pump energies, with a rapid increase in the peak intensity at pump energies above 48 μJ, and a sharp decrease in the line width to sub-nanometers with the increase in pump energy. This important feature indicates that the operating threshold of the prepared POF-based coherent RL is about 48 μJ.

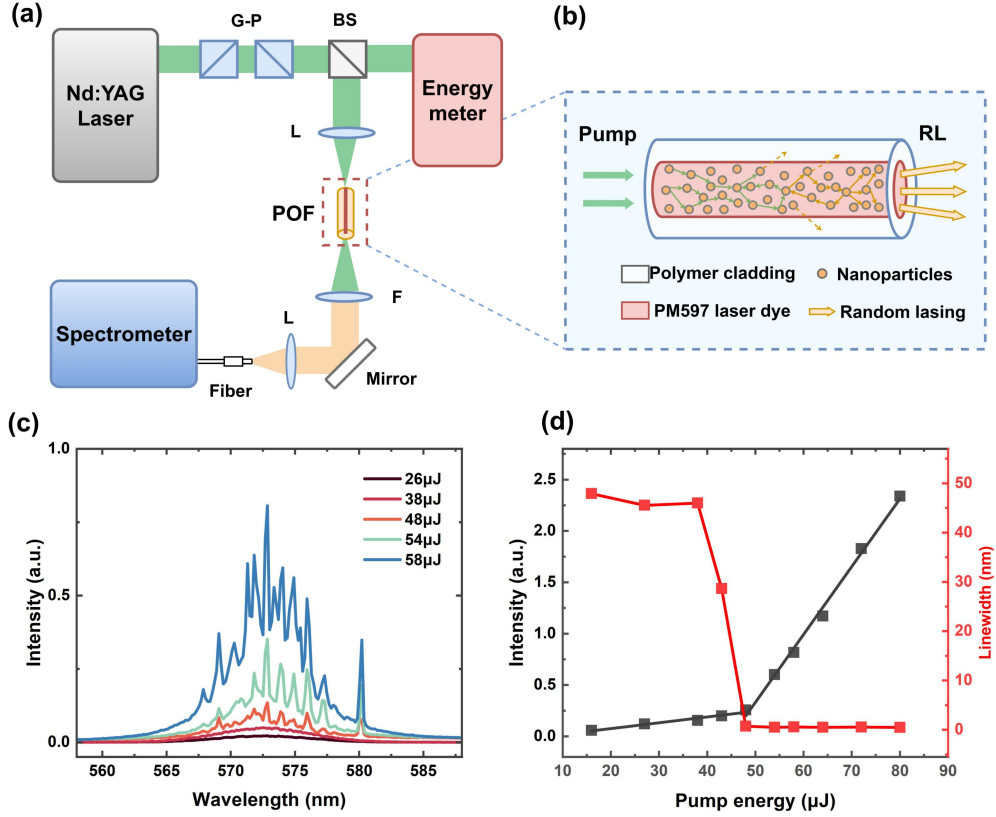

**Figure 1. Coherent random laser (RL) principle and output properties based on polymer optical fiber (POF).** (**a**) The experimental setup. G-P, Glan-Prism group; BS, beam splitter; L, lens; F, filter. (**b**) Schematic diagram of RL generation in POF. (**c**) Emission spectra of the sample at different pump energies. (**d**) Intensity and line width of the sample as a function of pump energy, with a threshold value of about 48 μJ.

## 2.2 Spectral data analysis

The collected spectral data of coherent RL are analyzed, with the pump energy stable at 80 μJ. The single-scan output spectra for T = 15.2 s, 21.1 s, 34.1 s, and 42.1 s are shown in **Figure 2a**. The spectra appear to have many spikes, and modes compete with each other, displaying different mode distributions at different moments. The long-term real-time spectral evolution of the coherent RL from 0 to 50 s is shown in Fig. 2b. The wavelength position of the peaks is fixed during the spectral evolution, which is due to coherent feedback induced by fixed scattering particles, but with random intensity. A real-time intensity scan at 572.8 nm for a single peak is shown in Fig. 2c, and the intensity fluctuates steadily and randomly. The autocorrelation function of Fig. 2c is analyzed (Fig. 2d), and there are no obvious spurious peaks, indicating low correlation of the individual wave peak channel. From the Pearson correlation coefficients of the valley (Fig. 2e) and peak (Fig. 2f) channels, the inter-modal correlations at different wavelength positions are compared. At the valley (Fig. 2e), the red regions are more covered, indicating a high intensity correlation. This may be due to the fact that the spectrum has a fixed envelope with relatively stable valleys. At the peaks (Fig. 2f), the gray regions are not complete, indicating that the intensities are not completely uncorrelated. The current peak is more or less influenced, to some degree, by the intensity of the other peaks. Fig. 2g shows the sampling locations of the peaks and valleys. The red points are the valleys, and the blue points are the peaks. The full spectral correlation heat map is in the supplementary information (Figure S1). By analyzing the coherent RL spectra, it is found that the coherent RL is temporally intensity uncorrelated in the individual wavelength channels, while there is some correlation between the wavelength components of the spectra.

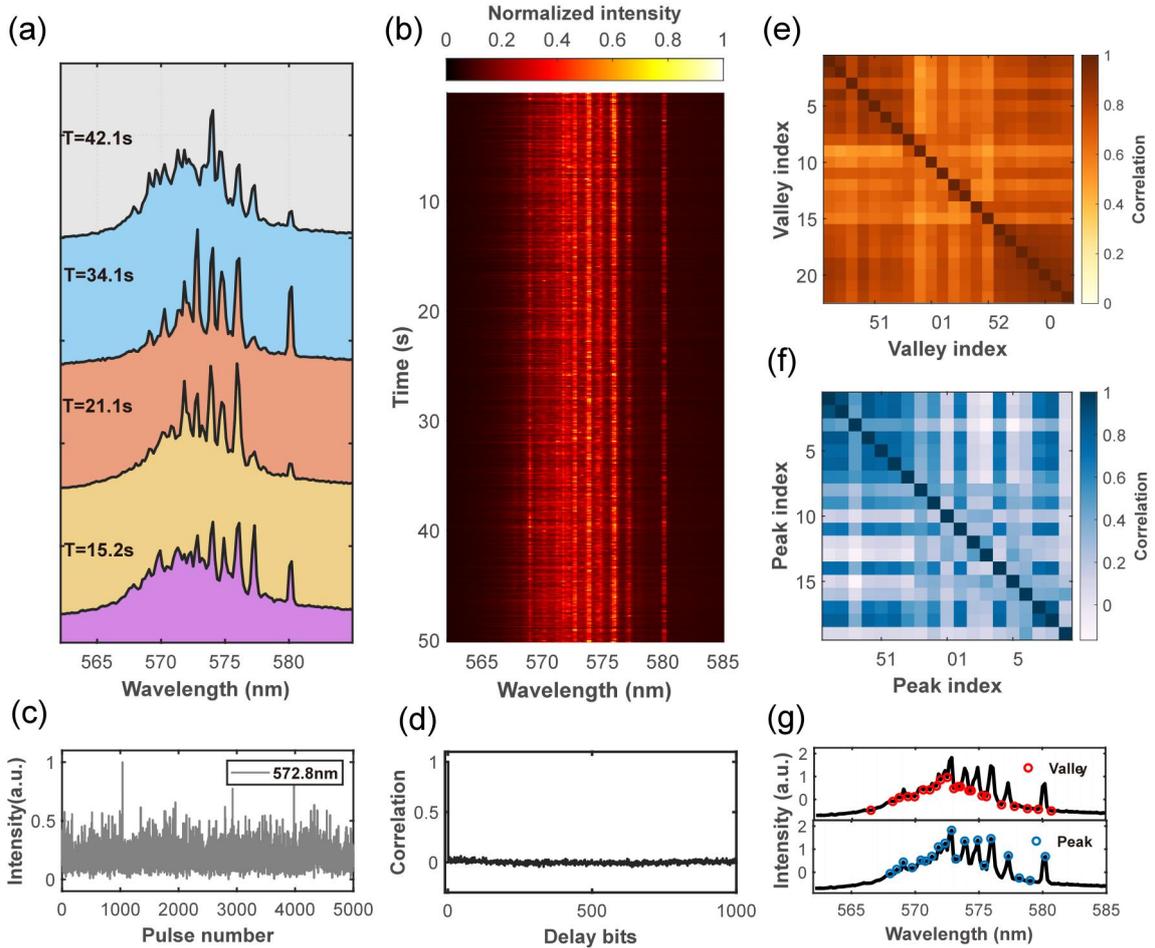

**Figure 2. Correlation analysis of coherent RL spectra.** (**a**) Single-scan spectra of the coherent RL at different moments, seen to comprise several narrowband features. (**b**) Long-term real-time spectral evolution obtained from consecutive scans. (**c**) Long-term real-time intensity variation at a wavelength of 572.8 nm. (**d**) The time autocorrelation function in (e) with a delay of 1000 bits. (**e** and **f**) Pearson correlation coefficient heatmaps of intensity

fluctuations at the valley and peaks, respectively. (**g**) The sampling locations of the peaks and valleys. The red points are the valleys, and the blue points are the peaks.

## 3. Temporal unpredictability of coherent RL intensity

### 3.1 Experimental setup

In order to verify that the temporal intensity fluctuations of coherent RL do not have a deterministic property similar to chaotic sequences, neural network models for time series prediction are necessary. The models are used to predict the subsequent RL intensity for each wavelength channel. The structural framework is shown in **Figure 3a**. The spectral history data of the first $t$ time steps are used as input. The wavelength components of the spectrum can be regarded as parallel time series. After that, the deep learning model will learn the temporal features in parallel and predict the spectrum at the $t+1$ time step. In this paper, the observation window T = 32. The previous $t$ time-step spectra are used as end-to-end inputs, and the $t+1$ time-step spectrum is used as an end-to-end output. Historical spectra are used to make predictions about spectra at subsequent moments. The goal is to determine whether the temporal intensity evolution of coherent RL is potentially deterministic.

### 3.2. Data pre-processing

Due to the different ranges of spectral data, the data is normalized for ease of training in order to train the model at a faster rate of gradient descent. The dataset is processed as continuous time series data by the sliding time step method. The dataset is from the spectral data under 5000 pulses, which is divided into two subsets: the training set and the test set. The data split ratio is 0.8/0.2, i.e., there are 4000 frames in the training set and 1000 frames in the test set. The test set is not involved in the training process, and the training set is not involved in the subsequent validation operations. In this paper, the spectra of the previous 32 time-steps are used as inputs for the next time step's spectra prediction.

### 3.3. Network models for time series prediction

Here, time series prediction models with excellent performance are necessary. In this paper, a variety of cutting-edge time series prediction models are used, including MLP-Attention (MLP-A) [41], GRU [44], CNN-LSTM-Attention (CLA) [42][43], [45], and Transformer [46] models. These models perform well in processing complex data and capturing time series features. In addition, pseudo-random sequences (PRS) with determinism are added as a control group for prediction to study the difference in randomness between with RL. The PRS is kept the same size as the spectral dataset and the model training parameters are identical. The generation of the PRS is detailed in Method. More details of the model are in the supplementary information.

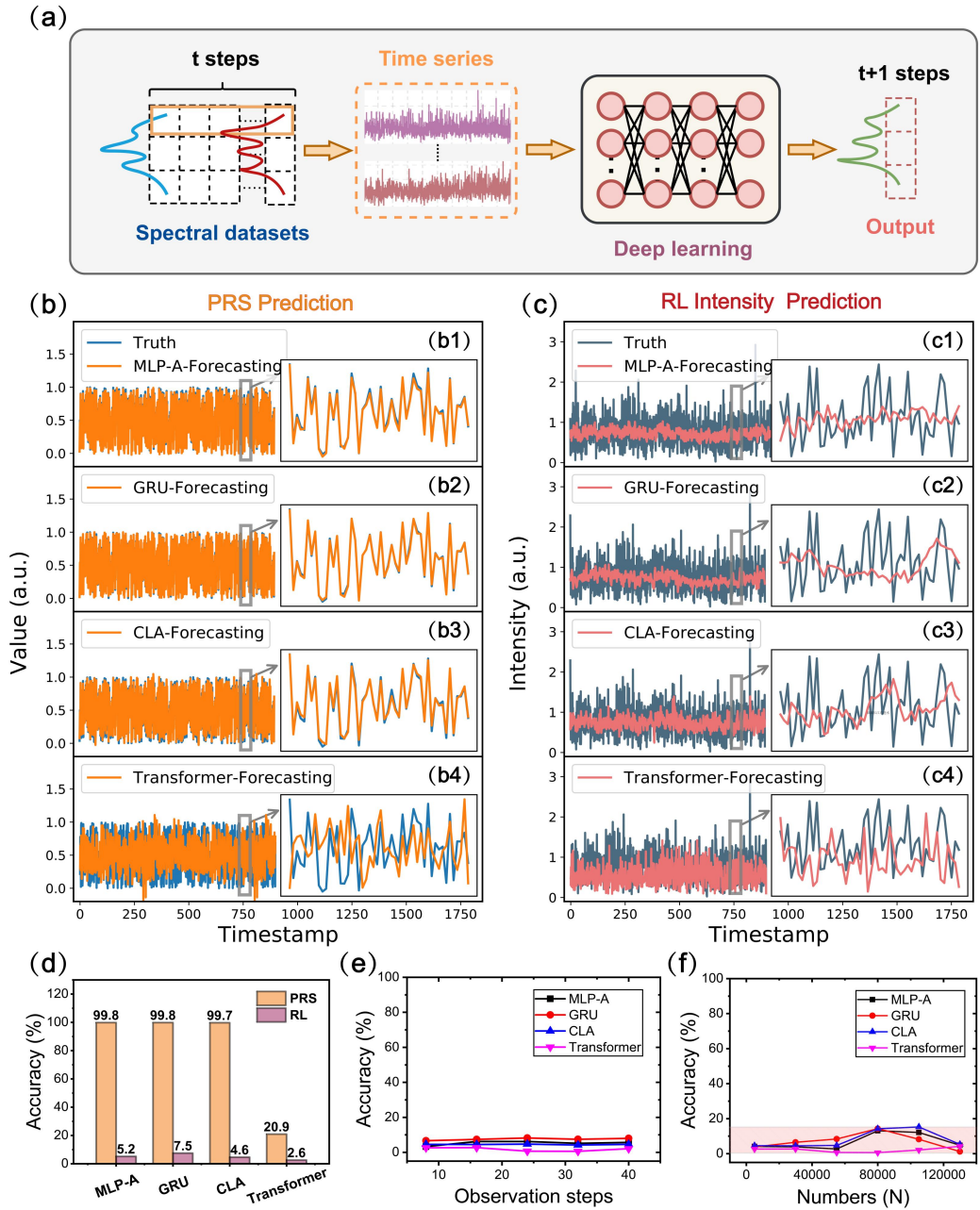

**Figure 3. Predicting the temporal evolution of RL spectra based on multiple time series prediction models.** (**a**) Schematic of spectral time series prediction. (**b**) PRS and (**c**) RL prediction performance of MLP-A, GRU, CLA and Transformer models, respectively. (**d**) Prediction accuracy of different prediction models for PRS and RL. (**e**) Prediction accuracy of model at different observation steps. (**f**) Prediction accuracy of RL intensity with dataset expanding for different prediction models.

### 3.4. Forecasting performance

After training, the predicted results in 1000 test stamps are obtained. In Fig. 3b the prediction performance of MLP-A (b1, b2), GRU (b3, b4), CLA (b5,b6), and the Transformer model (b7, b8) are shown for the PRS (predictable) versus RL, respectively. The insets show localized prediction details. None of the four models can successfully learn the time-intensity evolution of RL. the prediction performance of RL is all worse than that of PRS, indicating that RL is more unpredictable. The prediction accuracy results are shown in Fig. 3c. The MLP-A, GRU, CLA, and Transformer models are all unable to predict RL with 5.2 %, 7.5 %, 4.6 %, and 2.6% accuracy, respectively. Among them, MLP-A, GRU, and CLA all successfully predicted PRS with 99.8, 99.8, and 99.7 % accuracy, respectively. Far more than the prediction results of RL, which also indicates that RL has more randomness compared to PRS. The Transformer model, though, does not succed in predicting PRS with 20.9 % accuracy, which is mainly due to the fact that Transformer lacks the ability to memorize temporal information (even in the presence of positional encoding). However, the prediction accuracy is still higher than RL. The accuracy is relatively smooth when the historical observation time step is at 4, 8, 16, 32 and 40 (Fig. 3h). The results indicate that the increase in observation time step is not effective in reducing the prediction error, and the historical time step provides a lower contribution. Fig. 3e exhibits the accuracy results of MLP-A, GRU, CLA, and Transformer trained from 5,000 to 130,000 spectral predictions. The prediction model consistently maintains a lower accuracy of 3.8 to 15.3 % as the dataset expands. The stable low accuracy indicates that even if the amount of data is increased, RL remains unattainable for prediction by existing AI models. RL can be used as a reliable source of entropy.

### 4. Inter-modal predictability of coherent RL spectra

#### 4.1. Data downgrading

In order to improve the training efficiency, the data needs to undergo dimension reduction before training. Due to the fixation of the scattering nanoparticles in the POF, the position of the peaks in the emission spectrum does not change from one shot to the next [53]. The most wavelength position information of the peaks has already been determined, so the DNN only needs to learn the correlation between the intensities of the peaks. As shown in **Figure 4a**, the wavelength dimension can be reduced by only sampling the peaks and valleys of the spectra, since the wavelength positions of the peaks and valleys remain stable. After sampling, the wavelength dimension of the data is reduced from 134 to 43 dimensions. The dataset is the spectral data under 5000 pulses, which is divided into two subsets: the training set and the test set. The data split ratio is 0.8/0.2, i.e., there are 4000 frames in the training set and 1000 frames in the test set. It is worth noting that the test set is not involved in the training process, and the training set is not involved in the subsequent validation operations.

### 4.2. Experimental setup

The correlation between the coherent RL spectral modes suggests that there may be some underlying correlation between the spectral components. To explore whether such inter-modal intensity correlations can be determined, DNN is used to learn this nonlinear relationship. The DNN attempts to predict the intensity of the missing peaks using the known partial RL spectra. The principle is shown in Fig. 4b. The intensity of a peak is erased from the spectrum and then recovered using only the residual peak. Prior to this, the DNN needs to be trained with a large amount of spectral data.

### 4.3. The DNN model

The DNN is designed to extract hidden features from other wavelength components and consists of 4 FC layers. Due to the different ranges of spectral data, the data is normalized for ease of training so that the model can be trained at a faster gradient descent rate. For example, the intensity values of the known 42-dimensional peaks and valleys are utilized to predict the unknown 1-peak intensity distribution. It is worth noting that the input data is only sourced from wavelength component data at the same moment in time, independent of the spectral data from the previous time step. More details of the model are in the supplementary information (Figure S3).

### 4.4. Forecasting performance

After training with the neutral network, the predicted results in 1000 test stamps are obtained. Figs. 4c and d show the forecasting results when 1-peak is predicted. The blue line is the ground truth, and the red line is forecasting. In Fig. 4c, the 1-peak prediction at a certain moment is highly accurate, and the inset shows a zoomed-in display of the predicted crest region. In Fig. 4d, the peak prediction for the 572.8 nm wavelength channel at 1000 steps in the test set fits extremely well. The magnified region is between 450 and 500 time steps. After the training of the DNN model, the intensity relationship between the unknown modes is predicted correctly. Figs. 3e and 3f show the corresponding results when 3-peaks are predicted. The forecasting performance is somewhat degraded, but still matches well. Figs. 4g and 4h show the loss of the model with the training epochs when predicting 1 and 3 peaks, respectively. The loss converges gradually with the training epochs, and eventually the mean square error (MSE) drops to about $10^{-3}$ for 1-peak and $10^{-1}$ for 3-peak, respectively.

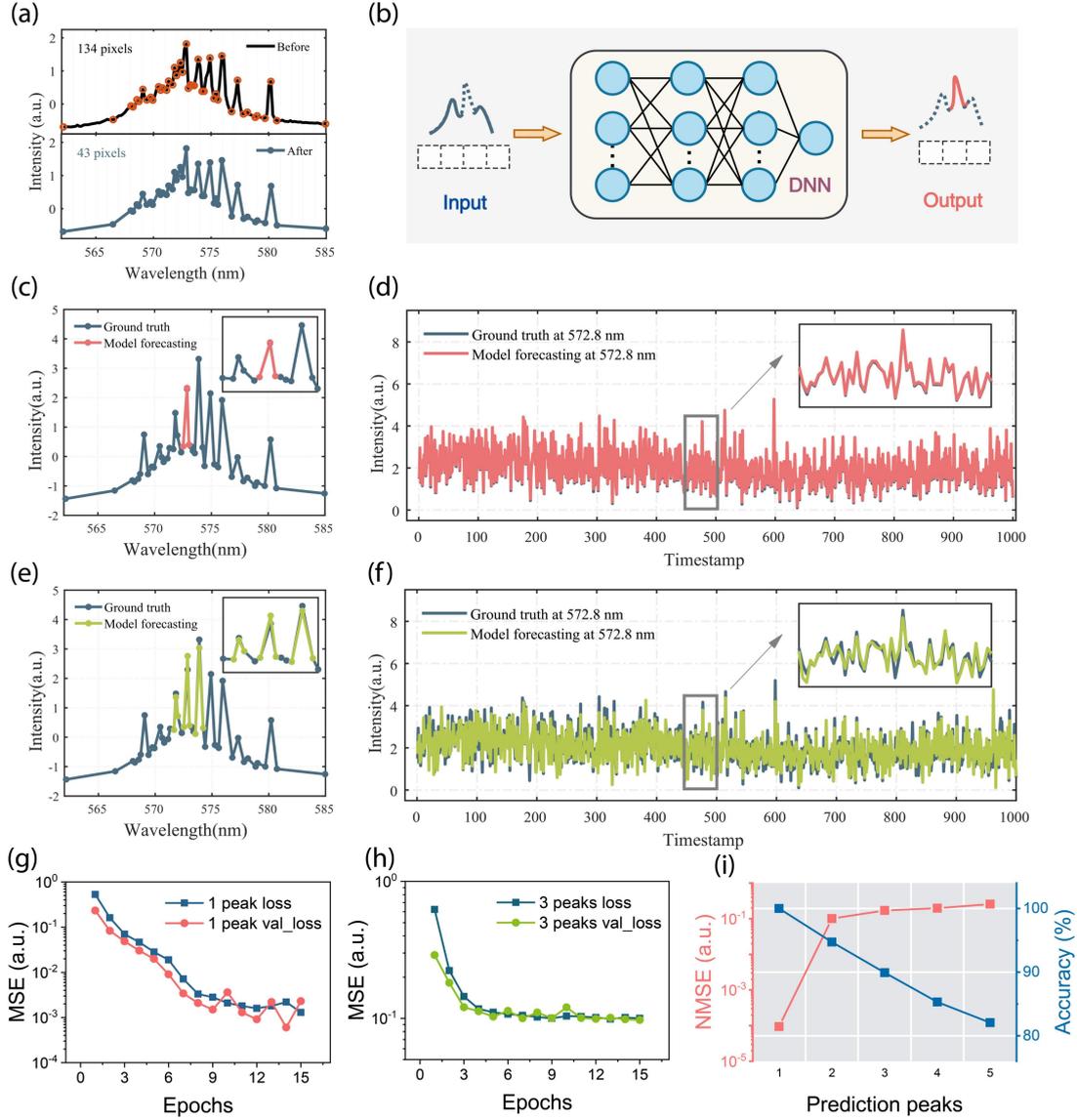

**Figure 4. The DNN model for coherent RL inter-mode prediction and forecasting performance.** (**a**) Spectral data dimension reduction. (**b**) Schematic diagram of RL unknown mode prediction using the DNN. (**c**) Forecasting scenarios of 1 peak. (**d**) Prediction performance of 1 peak in 1000 test stamps. (**e** and **f**) The corresponding results when 3-peaks are predicted. (**g** and **h**) Mean square errors (MSE) of 1 and 3 peaks with the training epochs, respectively. (**i**) Normalized mean square errors (NMSE) and accuracy of model predictions with different prediction peaks.

The DNN has learned effective features, and there is indeed a potential regularity in the mode distribution of coherent RL. The errors at predicted peaks of 1, 2, 3, 4, and 5 are compared in Fig. 4i. The normalized mean square errors (NMSE) are $10^{-4}$, $10^{-1}$, $1.7 \times 10^{-1}$, $2.0 \times 10^{-1}$, and $2.6 \times 10^{-1}$, respectively. The results show that the prediction error increases linearly with the increase in the number of predicted peaks, and there is no deterioration of super linearity. It is understandable that an increase in the number of predicted peaks leads to a decrease in the data of known peaks and a subsequent deterioration in the training results. Meanwhile, the accuracies in the test set are 99.99%, 94.73%, 89.93%, 85.30%, and 82.06%, respectively. The benchmark results indicate that the DNN model has relatively good results and robust performance for pattern competition relationships in coherent RL. Although RL exhibit unpredictable randomness in the time domain, a highly predictable determinism exists between the spectral components. RL possesses both random and deterministic properties.

## 5. Secure key generation and distribution based on random polymer fiber laser (RPFL)

### 5.1. System architecture

Based on the dual properties of randomness and determinism of coherent RL, an innovative secure key generation and distribution scheme is designed in this paper. The scheme cleverly utilizes the spectral properties of RLs. The system architecture is shown in **Figure 5**. Alice and Bob specify the temporal intensity at a certain wave peak of the spectra as the key. Alice acts as the key transmitter, and the random polymer fiber laser (RPFL) with multiple random spectra serves as the generator of the key. After the spectra pass through the spectroscopic device, the intensity of one of the peaks is separated. Alice converts the intensity information of the peaks into random bits and saves them as key A. The remaining spectral information is transmitted to Bob's end through free-space optics. At Bob's end, the peak and valley information of the incomplete spectrum is sampled and restored using a pre-trained DNN network. The restored peak information is converted into random bits and saved as key B. Key A and key B are identical, thus completing the key generation and distribution process. Additionally, we have included two attackers, Eve1 and Eve2. The attackers receive waveforms with more noise and some distortion. Assuming Eve1 does not have the same DNN for recovery, it can only sample from neighborhood peaks, converting them into random bits and obtaining an Error Key 1. Eve2 has the same DNN, but due to waveform distortion, the sampling positions will drift from the correct ones. After recovery and conversion into random bits through the DNN, it obtains an Error Key 2.

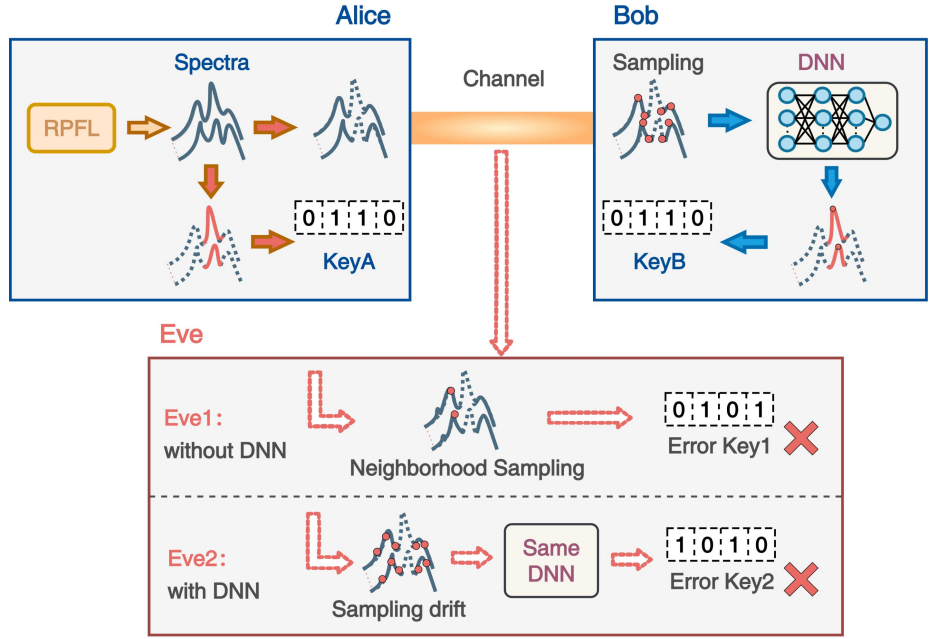

**Figure 5.** Architecture diagram of the key distribution system based on random polymer fiber laser (RPFL).

## 5.2. Key generation

The time series of each peak of the coherent RL are difficult to be predicted by AI models. This unpredictable intensity fluctuation is highly randomized, which can be used for random bit generation of keys. **Figure 6a** shows the processing of RL intensity of a single wave peak converted into random bits. In this case, the first 128 pulse time intensity sequence of the RPFL at 572.8 nm is shown in Fig. 6(a1). The corresponding amplitude probability distribution (a2) is not symmetrically distributed and has a statistical deviation. The self-delayed difference method is a method to eliminate the statistical bias. For intensity, the first order difference is defined as: where means the time difference between two neighboring data points. The intensity series after the self-delayed differencing process is shown in Fig. 6(a3). The probability distribution after differencing (a4) is symmetrically distributed along the left and right sides at amplitude 0. This ensures that the 01 distributions of the random bits after threshold judgment at 0 remain equal. The sequence of random bits after threshold judgment (a5) has only 01 signals. The corresponding amplitude probability distributions (a6) have equal 01 probability distributions of 0.5025 and 0.4975.

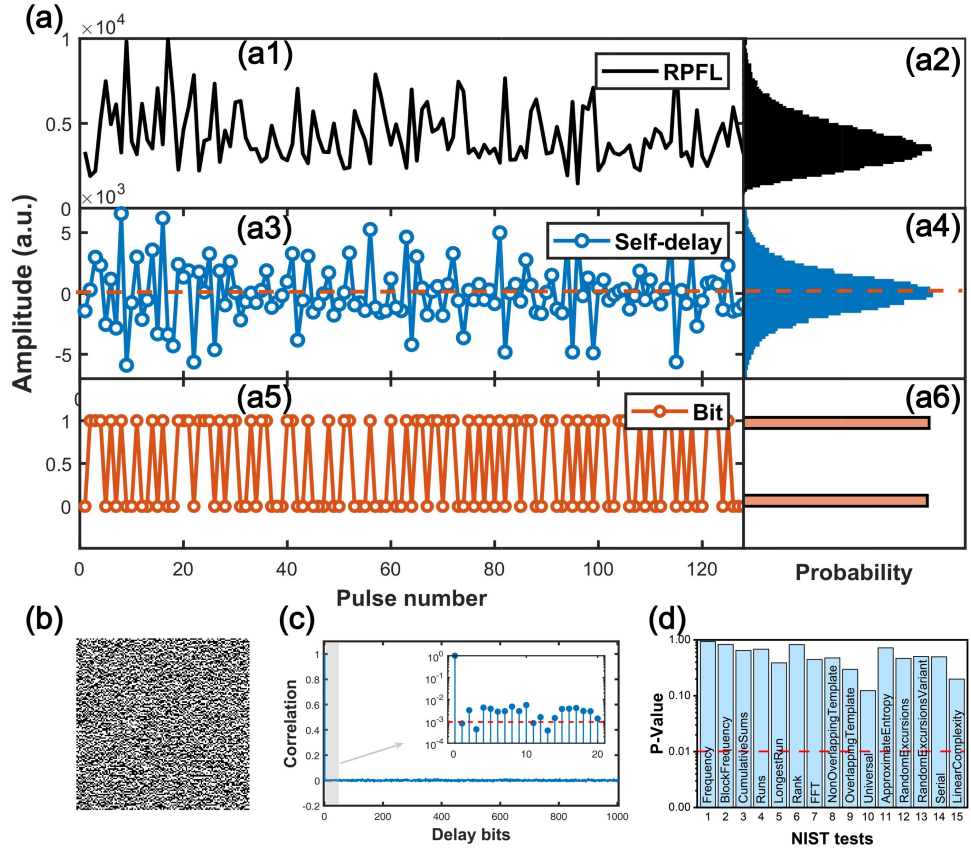

**Figure 6. Key generation based on temporal randomness of RL spectral components.** (**a**) Processing of the temporal intensity into random bits of individual peak. RPFL output at 572.8 nm, self-delayed difference-processed, threshold-judgmented time-intensity series (a1, a3, a5), with the corresponding amplitude probability distributions (a2, a4, a5), respectively. (**b**) 128×128 pixel bitmap. (**c**) Autocorrelation function (ACF) results for a random bit stream. Inset: details of ACF results for the first 20 delayed bits. (**d**) Results of 15 NIST tests.

The generated 128×128 pixel random bits are shown in Fig. 6b. No significant patterns or deviations are observed. The autocorrelation function (ACF) results are shown in Fig. 6c. The inset shows the details of the ACF results for the first 20 delayed bits. No significant peaks are observed along the delayed bits and the function value remains near $10^{-3}$, indicating that the temporal correlation of the random bits is extremely low. To further determine the randomness of the generated random bit stream, we use the state-of-the-art industry-standard statistical test suite NIST SP800-22 [54]. Fig. 6d shows the results of the NIST test suite. The passing criterion for each test is that the p-value should be greater than 0.01. The results of the 15 standard tests confirm that the generated random bit sequence passes all the NIST tests and can be considered as a random bit sequence with excellent performance.

### 5.3. Key distribution

After the key generation, the remaining spectra are transmitted through free space to Bob's end. Bob's end samples the information of the peaks and valleys of the residual spectra and implements the key recovery through the neural network. In **Figure 7**, the key recovery of Bob is simulated with the attack of Eve1 and Eve2. Eve channel is realized by adding Gaussian white noise. The key delivery wavelength channel is 572.8 nm. Eve1 has no DNN and therefore steals the intensity signal from the adjacent 571.6 nm channel. Eve2 has the DNN, but the presence of noise deviates the sampling position by plus or minus 0.2 nm. The correlation coefficient (CC) is used to describe the accuracy between the detected signal and the correct signal. The closer the CC is to 1, the higher the accuracy is. The waveforms of Alice and Bob, Alice and Eve1, and Alice and Eve2 are matched as shown in Fig. 7(a), (b), and (c), respectively. Among them, only Bob's waveforms match perfectly, Eve1 and Eve2 both have large deviations, and the waveforms have no obvious correspondence. It shows that only Bob recovers the correct key. Figs. 7(d), (e), and (f) show the pairwise correlation plots of Alice and Bob, Alice and Eve1, and Alice and Eve2, respectively. Among them, only Bob presents a high linear correlation. The CC reaches 0.9999, indicating the high accuracy of the recovered intensity information. On the other hand, the correlation graphs of both Eve1 and Eve2 are very scattered with CCs of 0.2545 and 0.1568, respectively, indicating that the accuracy of the recovered information is low and the correct key cannot be stolen. The excellent key distribution performance indicates that the scheme has high accuracy of key distribution and is resistant to stealing attacks.

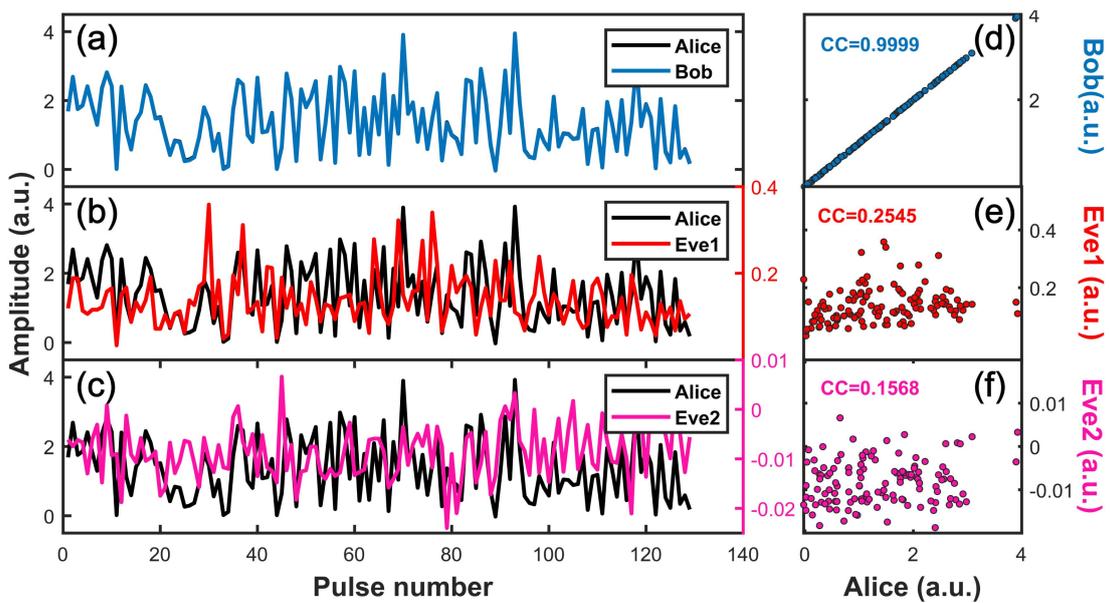

**Figure 7. Accuracy of receiver and eavesdropper after neural network key distribution.**

Signal similarities between (**a**) Alice and Bob, (**b**) Alice and Eve1, and (**c**) Alice and Eve2, respectively. Pairwise correlation plots of received signals from (**d**) Alice and Bob, (**e**) Alice and Eve1, and (**f**) Alice and Eve2, respectively.

A highly robust key distribution scheme can not only guarantee the security and reliability of key distribution, but also improve the flexibility of the system and reduce the maintenance cost. The robustness of the scheme is analyzed in **Figure 8**. Fig. 8(a) shows that Bob's Bit Error Rate (BER) remains stable (about $5\times10^{-3}$) over 2000 to 20,000 test sets with different number of predictions. The BERs of Eve1 and Eve2 are about 0.45, and the key distribution is always stable during the working period. Fig. 8(b) shows that Bob's prediction accuracy is as high as 99.99% under different power pumping (60~140 µJ, the pulse width of the laser in this paper is 6 ns, corresponding to an instantaneous power of 10~23.3 W), while the accuracy of Eve1 and Eve2 is only about 50%. Fig. 8(c) shows that the BER of Bob is lower than that of the Eve1 and Eve2 at different signal-to-noise ratios (SNR). Bob is $\sim10^{-2}$ at 50 dB. The BER of Eve1 stabilizes at 0.5, and the BER of Eve2 gradually stabilizes from 0.5 to 0.3. The scheme is noise-resistant, and it is difficult for the stealing party to obtain the correct information. Fig. 8(d) shows that the symbol error rate (SER) of Bob and Eve 1, 2. 128 bits are used as a code symbol. The details of the inset show that the SER of Eve 1, 2 is always 1, and Bob is 0.73 at 50dB. Error correcting codes (ECC) can continue to reduce the SER of the message transmission, and with the introduction of Reed-Solomon ECC [55], the SER at Bob's is significantly reduced (e.g., 64-ECC from 1 to $2.9\times10^{-2}$ at 30dB), which ensures the accurate transmission of the key. In contrast, Eve 1, 2 always have a SER of 100% when there is no ECC. The correct key cannot be obtained by Eve 1, 2.

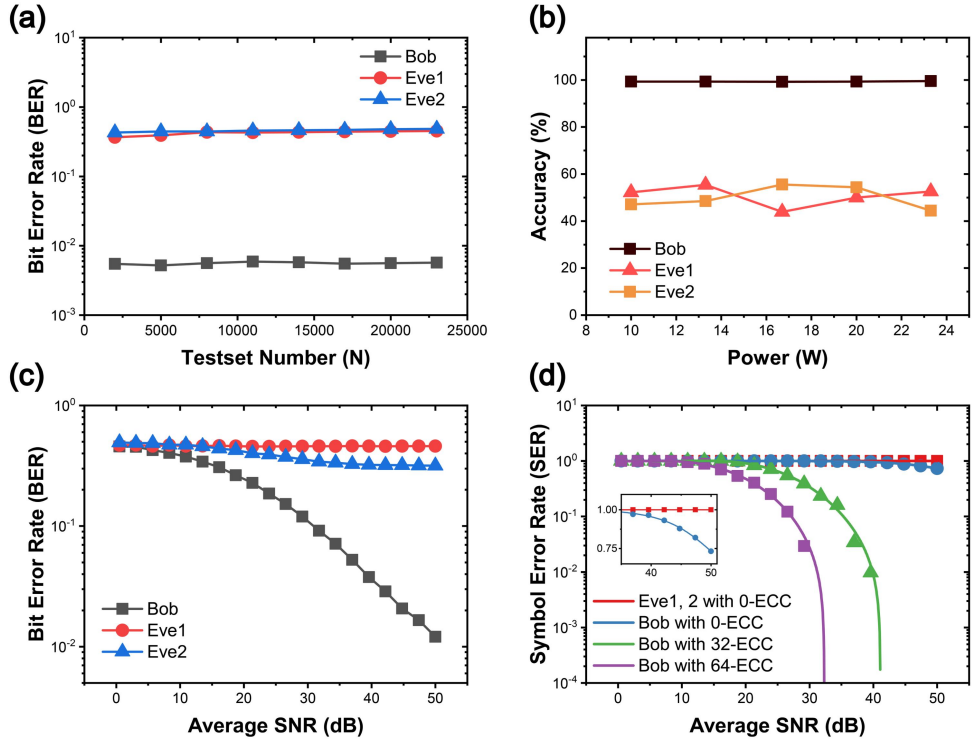

**Figure 8. Robustness analysis of key distribution.** (**a**) Bit error rate (BER) performance of Bob with Eve 1, 2 under different testset number. (**b**) Accuracy performance of Bob with Eve 1, 2 under different instantaneous pump powers. (**c**) BER performance of Bob with Eve 1, 2 for different signal-to-noise ratios (SNR). (**d**) Effect of error correcting code (ECC) on symbol error rate (SER) of Bob and Eve 1, 2 at different SNRs. 128 bits are used as a code symbol. The inset shows the BER details of Bob and Eve 1, 2.

## 6. Discussion

In summary, our study indicate that RLs possess both random and deterministic properties. The unique photonic property can be used to design SKGD schemes. RLs have a high degree of randomness as a source of physical entropy. A variety of cutting-edge time series prediction models (including MLP-A, GRU, CLA, and Transformer) are unable to accurately predict the output of RLs. This result further solidifies the conventional knowledge that RLs are highly random in the time domain. However, as the study progresses, we find that RLs are not completely inscrutable. Between the spectral wavelength components, the intensity hides a deterministic law that can be predicted. Using a DNN model, the complex nonlinear relationship between the inter-modal wavelength components of coherent random lasers is successfully learned. The intensity between wave peaks is predicted deterministically. Experimental results show that the prediction accuracy of a single peak is as high as 99.99%. Although the RL exhibits unpredictable randomness in the time domain, a highly predictable determinism exists between the spectral components. RL possesses the dual properties of randomness and determinism at the same time.

Based on this unique characteristic, we have designed an innovative SKGD scheme. The scheme cleverly utilizes the spectral properties of RLs, which not only generates keys efficiently, but also ensures the secure distribution of keys. Compared with the traditional scheme, this new scheme demonstrates significant advantages in terms of security, efficiency and feasibility, and provides a brand-new strategy in the field of secure communication. And it is expected to play an important role in the future of encrypted communication, internet of things security, cloud computing security, and many other fields. This study not only breaks through the traditional knowledge of RLs in theory, but also brings innovative solutions to the field of information security at the practical level. With further exploration, RLs are expected to become a powerful candidate technology in the field of information security, providing a solid guarantee for protecting our digital world.

Experimental conditions limit our sampling rate, therefore, the key generation rate is not considered in this paper. However, the rate can be improved by devices with faster pumping frequencies such as picosecond lasers or continuous wave (CW) lasers [56][57]. Based on the work of LechSznitko [20], Riccardo Sapienza [33], and Li [24] et al., the generation rate can be improved in the time domain. Pump pulse widths of a few nanoseconds can explore the temporal sampling rate limitations of coherent RL systems. Typical RL relaxation times lie in the range between 1 and 10 ps, which implies that modulation frequencies of random lasers can theoretically be realized in the range of 100 to 1000 GHz. Under CW laser conditions with the help of high-speed photodiodes, 540 Gbps can also be achieved [57].

## 7. Materials and methods

*Experimental setup*: The coherent RL emission is from POF doped with the laser dye Pyrromethene 597 (PM597, 0.14 wt%) and $Fe_3O_4$ particles (0.2 wt%) [25]. The experimental setup is shown in Fig. 1a. A Q-switched Nd: YAG laser that outputs a wavelength of 532 nm (pulse duration 6 ns, repetition rate 10 Hz) is used to end pump the POF (8 cm long) with a convex lens (L) and 532 nm optical filter (F). The pump pulse energy and polarization are controlled by a Glan-Prism (G-P) group. The BS splits the pump light equally into two beams; one beam pumps a coherent random laser to the RPFL and is collected by a fiber spectrometer (SP) (QE65PRO, ocean optics, resolution ~0.4 nm, integration time 100 ms). The other beam is used to connect an energy meter (VEGA, ophir) to synchronize the recording of pump energy. 5000 spectra generated at 100 ms intervals and pump energy data are used as the datasets. In our example, there are no missing values or outliers; each spectrum consists of 134 points, and the training data are the spectra of 43 points after dimensionality reduction.

*Network parameter setting*: The model frameworks are built on a desktop PC with a 2.5 GHz Intel i5-12400 processor and 16 GB of memory using the Keras library [58] and the Tensorflow backend [59]. The computationally efficient Adam Optimizer showed slightly better results than other candidates [60]. The input data is in the form of (*b, t, n*), where *b* is the number of samples required for one training round, *t* is the historical time step required for training, and *n* is the variable dimension to be predicted. In the time series prediction models, the input data is (64, 32, 134) and the output data is (64, 1, 134), i.e., the spectral data from the first 32 timesteps is used to make a prediction of the spectrum at the next time step. In the DNN model, the input data is (32, 1, 42) and the output data is (32, 1, 1), i.e., the intensity values of the known 40-dimensional peaks and valleys are utilized to predict the intensity distribution of the unknown 1 peak. The performance changes can be confirmed by adding or subtracting parameters. By grid searching on the values, the historical data needed for forecasting. The optimizer is Adam, and the dropout rate is 0.1. For the performance evaluation, mean square error (MSE) is used. The formulas are as follows:

$$MSE = \frac{1}{n}\sum_{i=1}^{n}(\hat{y}_i - y_i)^2 \qquad (1)$$

where $\hat{y}_i, y_i$ are the predicted and ground truth spectra, respectively. *n* is the number of spectra in the validation set.

*Pseudo-random sequences (PRS) generation:* PRS is deterministic, predictable random numbers. The classical Linear Congruential Generator (LCG) algorithm [61] is used to generate PRS. The formula is as follows: $Xn = aX_{n-1} + c(\mathrm{mod}M)$. Where *a*, *c*, *M* are constants set by the generator. In this paper, we take *a* = 16807, *c* = 1, *M* = $2^{12}$, $X_0$ = 12345. The sequence generation length is the same size as the spectral dataset, which is 5000×134.

**Acknowledgements**

We especially thank Professor Ernesto P. Raposo at the Universidade Federal de Pernambuco, Brazil, for great discussions and providing suggestions for improvement. The authors would like to acknowledge the financial support from the National Natural Science Foundation of China (Grant Nos. 12174002, 11874012, 11404087), Excellent Scientific Research and Innovation Team of Anhui Province (Grant No. 2022AH010003), The innovation project for the Returned Overseas Scholars of Anhui Province (Grant No. 2021LCX011), Hefei

Municipal Natural Science Foundation (No. HZR2401). The participation of ASLG in this work is supported by CNPq, CAPES and FACEPE, Brazilian Agencies, and National Institute of Photonics (Grant No CNPq 465.763/2014-6).

**Competing interests**

The authors declare no competing interests.